\documentclass{article}

\usepackage[numbers]{natbib}

    \usepackage[preprint]{neurips_2024}

\usepackage[T1]{fontenc}    
\usepackage{hyperref}       
\usepackage{url}            
\usepackage{booktabs}       
\usepackage{amsfonts}       
\usepackage{nicefrac}       
\usepackage{microtype}      
\usepackage{xcolor}         
\usepackage{amsmath} 
\usepackage{times}
\usepackage{latexsym}
\usepackage{tabularx}
\usepackage{multirow}
\usepackage{multicol}
\usepackage{graphicx}
\usepackage{standalone}
\usepackage[T1]{fontenc}
\usepackage{microtype}
\usepackage{inconsolata}
\newcounter{stage} 
\title{Reinforcement Learning for Enhanced Targeted Molecule Generation Via Language Models}

\author{%
Salma J. Ahmed \quad Emad A. Mohammed\\
Department of Physics and Computer Science\\
Wilfrid Laurier University\\
Waterloo, ON N2L 3C5 \\
\texttt{\{ahme3460,emohammed\}@\{mylaurier,wlu\}.ca}}

\begin{document}

\maketitle

\begin{abstract}
Developing new drugs is laborious and costly, demanding extensive time investment. In this paper, we introduce a de-novo drug design strategy, which harnesses the capabilities of language models to devise targeted drugs for specific proteins. Employing a Reinforcement Learning (RL) framework utilizing Proximal Policy Optimization (PPO), we refine the model to acquire a policy for generating drugs tailored to protein targets. The proposed method integrates a composite reward function, combining considerations of drug-target interaction and molecular validity. Following RL fine-tuning, the proposed method demonstrates promising outcomes, yielding notable improvements in molecular validity, interaction efficacy, and critical chemical properties, achieving 65.37 for Quantitative Estimation of Drug-likeness (QED), 321.55 for Molecular Weight (MW), and 4.47 for Octanol-Water Partition Coefficient (logP), respectively. Furthermore, out of the generated drugs, only 0.041\% do not exhibit novelty.
\end{abstract}

\section{Introduction}
\label{sec:introduction}
The journey from conceptualizing a potential drug to its market availability is lengthy and financially demanding \cite{hughes2011principles,berdigaliyev2020overview}. It must navigate through several critical phases to transform a chemical compound or entity into a viable treatment for human diseases. Initially, a specific molecular target (such as a DNA sequence or protein) associated with a disease must be identified \cite{deore2019stages,lurje2010egfr,ator2006overview}. This target serves as the focal point for drug development, offering the potential for therapeutic intervention. Before a drug can be administered to patients, it must undergo rigorous preclinical research trials. These trials, conducted either in vitro or in vivo, aim to evaluate the drug's safety profile and assess its effects on biological systems. This phase is pivotal in determining the drug's potential for therapeutic use and understanding its impact on the body.

The drug progresses to clinical research after successful pre-clinical trials, where its efficacy and safety are tested on human subjects. This phase involves carefully designed clinical trials conducted in multiple stages to gather comprehensive data on the drug's performance and potential side effects. Subsequently, the accumulated data undergoes thorough scrutiny during the FDA review process, where regulatory authorities assess the drug's safety, efficacy, and overall benefit-risk profile. This extensive journey, while essential for ensuring patient safety and the effectiveness of medications, poses significant challenges \cite{sinha2018drug,gediya2009promise}. The prolonged timeline and substantial financial investment associated with drug development contribute to the high costs and low success rates in discovering new drugs. Manual drug discovery development is a resource-intensive process. For instance, new drugs take an average of 12 and 7 years, respectively, from pre-clinical testing to approval, with development costs for drugs exceeding \$1 billion \cite{dimasi2016innovation,FDA2025,van2016drugs}. As a result, innovative approaches that streamline the drug discovery process and enhance efficiency are crucial for addressing unmet medical needs and advancing therapeutic interventions.

\begin{figure}
\centering
\includegraphics[width=0.9\textwidth]{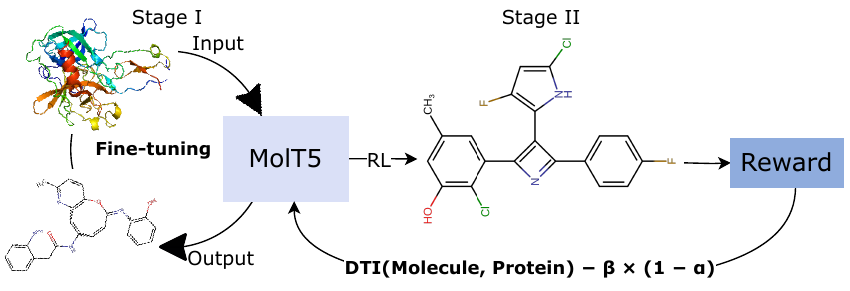}
\caption{MolT5, a Transformer-based model designed for molecular data, is initially fine-tuned (Stage I) for molecule generation given an input protein. Subsequently, the fine-tuned model undergoes reinforcement learning fine-tuning (Stage II) to enable targeted compound generation, using drug-target interaction and validity as rewards.}
\vspace{-10pt}
\label{pipe}
\end{figure}

With the rapid advancements in Deep Learning (DL), many researchers are exploring its application to address challenges in the drug discovery domain. These efforts span various areas, including Drug-Target Interaction (DTI), where models such as Transformers \cite{vaswani2017attention} and convolutional neural networks (CNN) \cite{wu2017introduction} are employed to identify interactions between known drugs and targets. This is a significant challenge in drug repositioning and can significantly accelerate the expensive and time-consuming drug development process \cite{wen2017deep,abbasi2021deep,you2019predicting}.

Another key area of focus is Protein-Protein Interactions (PPIs), which are crucial for understanding protein function, disease mechanisms, and therapy design. Many studies have developed deep learning models, such as autoencoders \cite{bank2023autoencoders}, to study sequence-based PPI prediction, achieving promising results \cite{sun2017sequence,hu2022deep,lee2023recent}.

Similarly, Drug-Drug Interaction (DDI) is critical for drug safety surveillance, enabling the effective and safe co-prescription of multiple drugs. Several studies have proposed using DL models to predict interactions in this domain \cite{deng2020multimodal,zhang2020deep,qiu2021comprehensive}.

Another critical area is Drug Re-purposing, where the same drug treats different diseases. To address this, several DL-based approaches, such as graph-based \cite{jiang2021could} and sequence-based models \cite{chen2023sequence}, have been proposed to re-purpose existing drugs for new therapies \cite{pan2022deep,hooshmand2021multimodal}.

De-Novo Drug Design is another key focus, generating novel compounds with desirable pharmacological and physicochemical properties. Various DL approaches, including recurrent neural networks, encoder-decoder architectures, reinforcement learning, and generative adversarial networks, have been applied to this task \cite{wang2022deep,krishnan2021accelerating,krishnan2021novo}.

These approaches highlight the potential of DL techniques to tackle complex challenges in drug discovery, offering promising avenues for identifying and developing novel therapeutics. This underscores the significant potential of employing deep learning methodologies to expedite the discovery of drug candidates and enhance our ability to combat emerging diseases.

In this paper, our primary focus is leveraging advancements in language models and reinforcement learning techniques to facilitate drug discovery through an approach for De-Novo drug design. Our main objective is to develop a generative model capable of taking a molecular target (protein) as input and using this model to generate novel molecules (drugs) specifically designed to treat or mitigate the effects of this protein within the body. The quality of the generated drugs can be evaluated by measuring their chemical properties to assess solubility and permeability, following \cite{lipinski1997experimental}.

The rest of the paper is organized as follows: Section II provides an overview of the related literature. Section III discusses the components of the proposed methodology. Section IV describes the experimental setup and presents the results. Section V offers an in-depth analysis and discussion of the findings. Finally, Section VI concludes the paper by highlighting key insights and directions for future research.

\section{Related Work}
Numerous approaches documented in the literature address the challenge of de novo drug design, given its significant potential impact when addressed effectively.

In \cite{popova2018deep}, the authors introduced ReLeaSE, a technique for producing novel targeted chemical compounds. This method integrated generative and predictive deep neural networks, where the generative model was trained to generate new chemical compounds while the predictive model forecasts desired properties. These models are trained independently and combined using a reinforcement learning approach to guide the generation process. Reinvent the Reinforcement Learning (RL) approach described by \cite{blaschke2020reinvent} for generating novel molecules that interact with specific targets by using RL to steer the generative model. This approach employs two Recurrent Neural Networks (RNNs) in an actor-critic setup. The critic, acting as the prior RNN, retains previous knowledge of the SMILES representation. The actor, or agent, is either an identical copy of the prior or a modified version that has undergone some initial training. The agent selects actions by sampling a batch of SMILES (S), which are then evaluated by the prior RNN and scored using a predefined scoring function. The method in \cite{gupta2018generative} introduced a generative RNN-LSTM model for drug synthesis. Initially, the model undergoes training on a dataset of molecular structures to grasp the syntax and patterns inherent in these representations. Subsequently, the RNN-LSTM model undergoes fine-tuning to skew predictions towards specific molecular targets. This adjustment is achieved by leveraging insights gained from the initial training to tailor the model to particular target molecules. Training on a generator RNN model conducted by \cite{zhang2023universal} using molecular data to familiarize it with the syntax of SMILES for molecular representation, enabling the design of novel and effective small molecules. Subsequently, they developed a drug-target interaction model to serve as a reward in a reinforcement learning framework. This model was employed to steer the generation of the RNN model towards specific properties or targeted molecules. A multi-objective approach was introduced by \cite{monteiro2023fsm} for drug synthesis, emphasizing properties and selectivity tailored to biological targets. Their method employed a transformer decoder to create drugs and a transformer encoder to forecast desired properties and refine learning through a feedback loop.

\section{Methods}
In this section, we thoroughly examine the methodology components, spanning from the dataset utilized to the generative model and the elements of the reinforcement learning paradigm.

\subsection{Dataset}
In all experiments and components of the proposed method, we use BindingDB \cite{liu2007bindingdb}, a publicly available database containing binding affinities of protein-ligand complexes. The dataset includes target proteins that serve as drug targets, with their corresponding structural information obtained from the Protein Data Bank \cite{bank1971protein}. It contains about 2,993,668 binding complexes data utilizing 9,499 proteins and over 1,301,732 molecules. This dataset is instrumental in achieving our objectives. Furthermore, a filtering criterion was applied to exclude complexes with protein amino acid lengths exceeding 500, primarily due to computational resource limitations.

\subsection{Generative Model}
\begin{figure}[ht]
\centering
\includegraphics[width=0.8\textwidth]{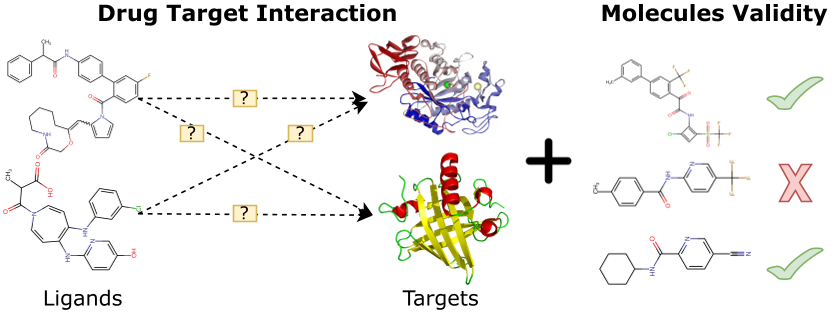}
\caption{The lefthand side depicts the methods used for reward calculation, focusing on drug-target interaction, while the righthand side illustrates the evaluation of molecule validity.}
\vspace{-10pt}
\label{RewardFig}
\end{figure}
An essential aspect of the methodology involves utilizing a generative model, which is vital for generating molecular compounds based on a given protein. We employed MolT5 \cite{edwards2022translation}, a self-supervised learning framework built upon an encoder-decoder transformer architecture \cite{vaswani2017attention}. MolT5 is pre-trained on unlabeled molecule compound strings and natural language text. Subsequently, the model underwent fine-tuning on two distinct tasks: molecule captioning, where the model receives a molecule string prompt and aims to generate a caption. In contrast, in the second task, text-based de novo molecule generation, the model generates a molecule string based on a provided textual description. Encouraging outcomes were achieved on both tasks, as stated by the authors, motivating the adoption of the model for our objective. Given a protein amino acid, the model should generate a targeted molecule drug. We utilize the base version of MolT5 (molt5-base) and conduct the first stage of the fine-tuning employing the protein-ligand complexes sourced from BindingDB to enhance the model's knowledge for our specific task, as depicted in Stage I of Figure \ref{pipe}, where the model takes a protein as input and generates a molecule, which is then compared with the true molecule in the protein-ligand complex.


\subsection{Drug-Target Interaction}
We aim to enable the generative model \cite{edwards2022translation} to generate molecule compounds tailored to specific proteins. To assess the activity of these molecules, we integrated a Drug Target Interaction (DTI) model using DeepPurpose \cite{huang2020deeppurpose}, a PyTorch toolkit for molecular modelling and prediction. It provides encodings for molecules and protein sequences fed into a multi-layer perceptron (MLP) decoder to predict binding scores. We tested two encoding methods: CNNs for SMILES strings and protein sequences and a CNN for protein sequences combined with transformer encoders for substructure fingerprints. Focusing on IC50 scores, we trained the models using BindingDB data and adopted an ensemble learning approach, merging predictions from both models. 


\subsection{Reinforcement Learning Setup} 
We used the Proximal Policy Optimization (PPO) algorithm \cite{schulman2017proximal}, a widely utilized method for training policies in reinforcement learning tasks \cite{yang2024crowdsourcing,das2024proximal,yau2024proximal,klein2024optimizing}, within our Reinforcement Learning (RL) framework. Our goal was to use this algorithm to guide the model in learning a policy where, upon receiving a protein amino acid sequence as input, the model would generate a molecule compound (drug) tailored to the specific protein, as shown in Stage II of Figure \ref{pipe}. We used the molt5-base model, which we had previously fine-tuned, as the starting point. We then fine-tuned it a second time using an RL policy. Reward selection is crucial in RL, as it acts as the feedback signal given to an agent, reflecting its actions and the state of the environment. It measures the agent's performance and indicates its effectiveness in completing tasks. Therefore, we adopted a composite reward strategy, as illustrated in Figure \ref{RewardFig}, to achieve the desired level of performance. The rewards used in our approach are as follows:

\begin{itemize}
    \item \textbf{DTI:} We integrated the DTI model into our RL methodology to serve as the reward mechanism. This decision aligned with our objective of training the model to generate molecules tailored to specific proteins. Consequently, when the model produces a molecule compound unrelated to the protein target, it receives a low reward score. Conversely, it gets a higher score when the generated molecule is targeted to the protein.
    
    \vspace{6pt}

    \item \textbf{Validity:} In addition to targeting specific proteins, we aim for the generated molecule to meet chemical validity criteria, ensuring its viability as a potential drug. Hence, we incorporated validity as the second reward within the RL paradigm. To assess validity, we employed the rdkit toolkit \cite{landrum2013rdkit} to determine whether the generated molecule complies with established chemical standards. 
\end{itemize}
\vspace{-1pt}
\begin{gather}
Reward = DTI( M_{G},P) - \beta \times ( 1 - \alpha ) \label{eq1}\\
\alpha = Validity(M_{G}) \in \{0, 1\}\label{eq2}
\end{gather}

\begin{figure}[ht]
\centering
\includegraphics[width=0.45\textwidth]{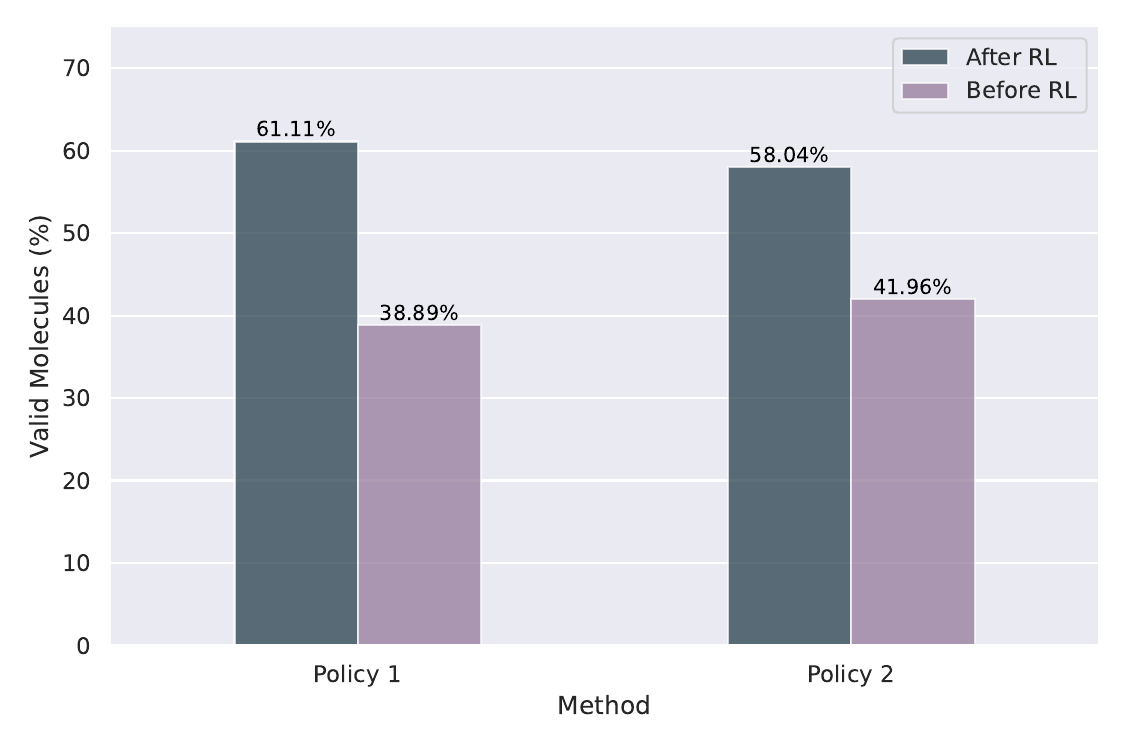}
\caption{The percentage of valid molecules before and after fine-tuning a model with reinforcement learning using two policies. Policy 1 combines Drug-Target Interaction (DTI) and molecule validity for rewards, while policy 2 only considers molecule validity. This evaluation assesses the method's effectiveness and adaptability to different objectives and policies.}
\vspace{-12pt}
\label{ApproachEfficacyFigure}
\end{figure}

Equation \ref{eq1} delineates the reward computation process. Initially, it computes the interaction between the protein ($P$) and the generated molecule ($M_{G}$) via the fine-tuned Drug Target Interaction model, denoted as $DTI$. Subsequently, the validity of the molecule is evaluated, with $\beta$ representing the penalty applied when the molecule is invalid ( $\beta = 0.30$). When the molecule is valid, indicated by $\alpha = 1$, the penalty term effectively reduces to zero. Equation \ref{eq2} elucidates the definition of $\alpha$, a binary variable with either 0 or 1 values, signifying the generated molecule's validity status.


\section{Experimental Setup and Results}

This section details each experimental procedure and its outcomes. Through systematic refinement of each component, we aimed to optimize performance and achieve superior results.

\subsection{Fine-Tuning Stage \textit{\Roman{stage}}}
Our initial fine-tuning stage focused on configuring the generative model to produce molecular compounds when provided with a protein amino acid sequence as input. This step represents a critical component of our experimental pipeline, as illustrated in Stage I of Figure \ref{pipe}. The model is fine-tuned on BindingDB complexes and evaluated using the Bilingual Evaluation Understudy (BLEU) metric \cite{papineni2002bleu}, which measures the similarity between the generated molecule and the actual molecule in the protein-ligand complexes, with the protein serving as input. Following this fine-tuning process, the model demonstrated the ability to create molecules based on protein inputs. However, our objectives extend beyond simple molecule generation; we aim to produce targeted molecules tailored to each protein. We conduct additional RL fine-tuning iterations to enable the model to learn a more refined policy.

\stepcounter{stage}
\subsection{Fine-tuning Stage \textit{\Roman{stage}}}
We employed the Transformer Reinforcement Learning (TRL) library \cite{vonwerra2022trl} for the RL fine-tuning. This library, built upon the transformers framework, is designed explicitly for refining transformer models through various techniques, including supervised fine-tuning (SFT), Reward Modeling (RM), Direct Preference Optimization (DPO), and Proximal Policy Optimization (PPO). We opted for the PPO method to fine-tune our model due to its recognized efficiency and effectiveness in training complex policies. PPO stands out for its ability to learn with fewer samples than alternative methods. Moreover, its utilization of a trust region optimization approach helps prevent drastic policy changes, ensuring stable learning dynamics. During experimentation, we explored different values for the policy optimization parameters, such as $top_k$ and $top_p$. Our analysis revealed that the most favourable results were achieved with $top_k$ set to 50 and $top_p$ to 0.95.
\begin{figure*}
\centering
\includegraphics[width=0.93\textwidth]{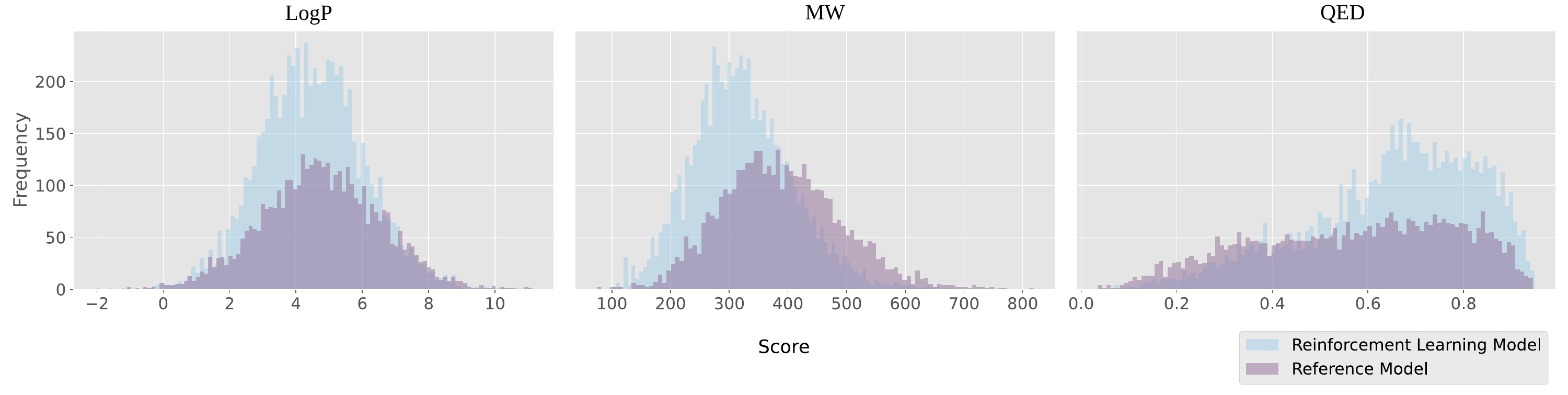}
\caption{This figure illustrates the distributions of the chemical properties of the generated molecules, including Octanol-Water Partition Coefficient (log P), Molecular Weight, and Quantitative Estimation of Drug-likeness (QED), both before and after fine-tuning the model with Reinforcement Learning.}
\label{ChemicalPropertiesFigure}
\end{figure*}

\vspace{7pt}

\begin{table}
\centering
\resizebox{10.0cm}{!}{%
\begin{tabular}{lcccc}
\toprule
\textbf{Chemical Properties}& 
\textbf{Before$_{RL}$} & \textbf{After$_{RL}$}\\
\midrule
Quantitative Estimation of Drug-likeness (QED) & 0.5705 & 0.6537\\
\midrule
Molecular Weight (MW) & 387.84 &  321.55\\
\midrule
Octanol-Water Partition Coefficient (logP) & 4.75 & 4.47\\
\bottomrule

\end{tabular}%
}
 \vspace{0.2cm} 
\caption{The resulting values of analyzing the chemical properties of the generated molecules both before and after fine-tuning the generative model with reinforcement learning.}
\vspace{-11pt}
\label{chemicalPropertiesTable}
\end{table}
\subsection{Reward Optimization}
The reward utilized for fine-tuning the generative model in RL consisted of two key components: drug-target interaction and the validity assessment of the generated molecule. We performed optimization to make sure their building and performance were good.
\vspace{4pt}

\textbf{DTI Enhancements}: The Mean Squared Error results for the CNN-CNN and the CNN-Transformer models were 0.004 and 0.006, respectively. We first experimented with using individual models in the RL framework and monitored the policy learning. Then, we combined two models using an ensemble learning strategy. We found that the optimal approach was to merge the predicted affinities of the two models, as shown in Equation \ref{eq3}. In this equation, $\textit{C}$ represents CNN, $\textit{T}$ represents Transformer, and $\textit{P}$ represents predictions. We assigned weights ($\textit{C}_w$ = 0.25 and $\textit{T}_w$ = 0.75) to each model's predictions to reflect their influence on the final affinity score, as these values were determined to yield the best performance. This fusion technique allowed us to leverage the strengths of both models, resulting in more robust and efficient policy optimization.

\begin{equation}
Affinity = (P_{C} \times C_{w} )+(P_{T} \times T_{w})\label{eq3}
\end{equation}
\vspace{0.1pt}

\textbf{Validity Integration}: Initially, our goal was for the model to generate targeted molecular compounds tailored exclusively to a specific protein. However, we analyzed the results and found that some of the generated molecules were chemically invalid. As a result, we introduced a new factor into the reward calculation: the validity of the generated compounds. This change enabled us to optimize the generative model to learn a single policy and develop a hybrid or mixed policy. This improvement ensures a more comprehensive optimization approach, considering the generated compounds' efficacy and chemical validity. The selection of the ($\beta$) value in equation \ref{eq1}, which represents the penalty applied to the reward for invalid molecules, was determined through trial and error. As we fine-tuned the model and monitored its learning and optimization, we experimented with values ranging from 0.1 to 0.7. After evaluation, we found that 0.3 produced the best results. This iterative approach ensured the penalty value was carefully adjusted to balance penalizing invalid molecules while maintaining effective model optimization. In Figure \ref{ApproachEfficacyFigure} for Policy 1, we show the percentage of valid molecules generated before and after applying RL fine-tuning with the combined reward.

\subsection{Model Adaptation to a Validity-only Policy}
To ascertain the effectiveness of our methodology, we undertook experiments to refine the model's focus solely on generating valid molecules. This involved modifying the reward calculation to prioritize validity without drug-target interaction. The notable optimization observed in the model's performance under this refined policy, as depicted in Figure \ref{ApproachEfficacyFigure} (policy 2), underscores the efficacy of this approach. These experiments offer compelling evidence of our methodology's ability to generate valid molecular compounds and highlight its adaptability to diverse learning policies.
\begin{figure*}
\centering
\includegraphics[width=0.7\textwidth]{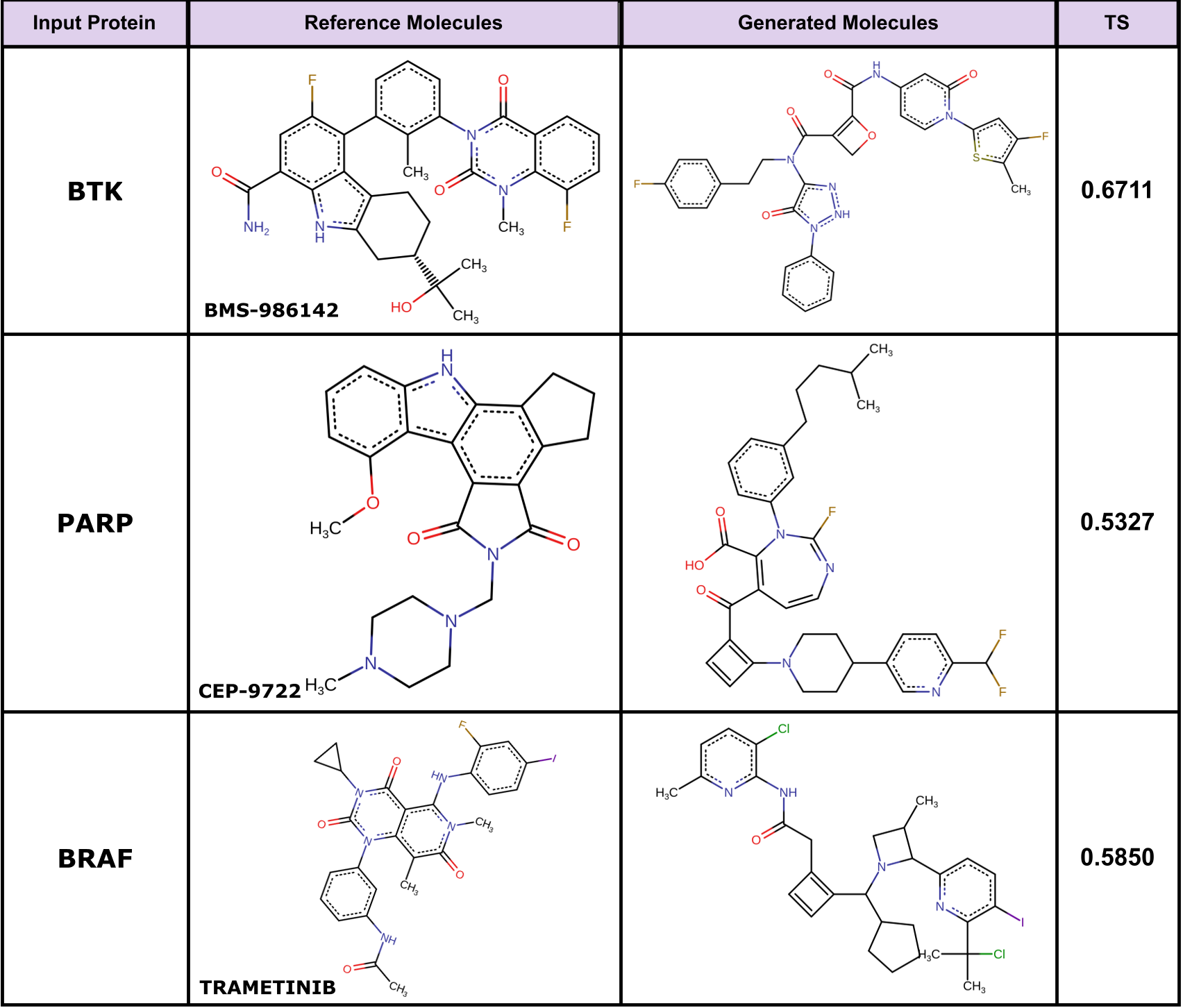}
\caption{Illustrative instances of the molecules generated through inputting different proteins into the model, juxtaposed with samples of protein inhibitors and the Tanimoto similarity ($TS$) between the generated and inhibitor compounds.}
\label{BigTable}
\end{figure*}

\section{Results Analysis and Discussion}
In this section, we delve into the comprehensive analysis of the proposed method, which includes assessing the chemical properties of the generated compounds and examining their interactions with proteins.

\subsection{Chemical Properties}
To evaluate the efficacy of our proposed approach, we analyzed the chemical properties of the generated molecules both before and after fine-tuning the generative model with reinforcement learning. We assessed a range of widely recognized molecular properties or descriptors commonly employed in drug discovery and computational chemistry, including Quantitative Estimation of Drug-likeness (QED) \cite{bickerton2012quantifying}, Molecular Weight (MW), and Octanol-Water Partition Coefficient (logP) \cite{lipinski1997experimental}. These metrics are typically utilized to gauge the potential of a compound to progress into a viable drug candidate.

\vspace{0.9pt}

\textbf{Quantitative Estimation of Drug-likeness (QED)}: This metric is used to evaluate the drug-likeness of chemical compounds or molecules, with values closer to 1 being better \cite{bickerton2012quantifying}. Integrating various molecular descriptors provides a numerical estimation of the likelihood that a molecule possesses favourable drug properties. We subjected the generative model to the same set of proteins to assess the QED of the generated molecules, both before and after fine-tuning with reinforcement learning. The results showed that the mean QED of the generated molecules was 0.5705 before RL fine-tuning and increased to 0.6537 after RL fine-tuning, as shown in Table \ref{chemicalPropertiesTable}. This indicates that the molecules generated after RL fine-tuning exhibit improved drug-like properties. 

\vspace{3pt}

\textbf{Molecular Weight (MW)}: This characteristic is crucial for understanding a compound's pharmacokinetics, formulation, and toxicity behaviour. It represents the sum of the atomic weights of all atoms within a molecule. Our evaluation included MW values of all generated molecules, conducted before and after RL fine-tuning, using the same set of proteins for assessment. Our analysis revealed that the mean MW value was 387.84 before fine-tuning, which decreased to 321.55 after RL fine-tuning, as shown in Table \ref{chemicalPropertiesTable}. This reduction in molecular weight offers potential advantages, as compounds with lower MW are often more easily absorbed and metabolized and may exhibit reduced complexity, facilitating more straightforward and accessible synthesis pathways. Desired values for molecular weight typically range from 250 to 500 g/mol for optimal drug-like properties \cite{lipinski1997experimental}.

\vspace{3pt}
\textbf{Octanol-Water Partition Coefficient (logP)}: Solubility and permeability are critical for predicting essential drug properties such as absorption, distribution, metabolism, and excretion (ADME). Therefore, logP serves as a key metric to assess the partitioning behaviour of a compound between an organic solvent and water. Our investigation revealed that the mean logP value was 4.7539 before fine-tuning, which decreased to 4.4766 after RL fine-tuning, as shown in Table \ref{chemicalPropertiesTable}. A logP value of 5 or less is often considered optimal for drug candidates, in line with Lipinski's Rule of Five for oral bioavailability \cite{lipinski2001cas}. We further analyzed the percentage of compounds with logP values less than or equal to 5, revealing figures of 62.972\% after RL fine-tuning and 56.106\% before RL fine-tuning. These results suggest that, after RL fine-tuning, a larger proportion of the generated compounds exhibited logP values within the optimal range, indicating favourable solubility and permeability characteristics. Thus, the fine-tuning process positively influenced the fundamental properties necessary for effective drug development.

\vspace{3pt}
As depicted in Figure \ref{ChemicalPropertiesFigure}, the distribution of the chemical properties of the generated molecules, including Octanol-Water Partition Coefficient (logP), Molecular Weight (MW), and Quantitative Estimation of Drug-likeness (QED), highlights the significant improvements and effectiveness achieved through the Reinforcement Learning fine-tuning of the model. This enhancement results in the molecules exhibiting the desired and optimal chemical properties.

\begin{figure}[ht]
\centering
\includegraphics[width=0.7\textwidth]{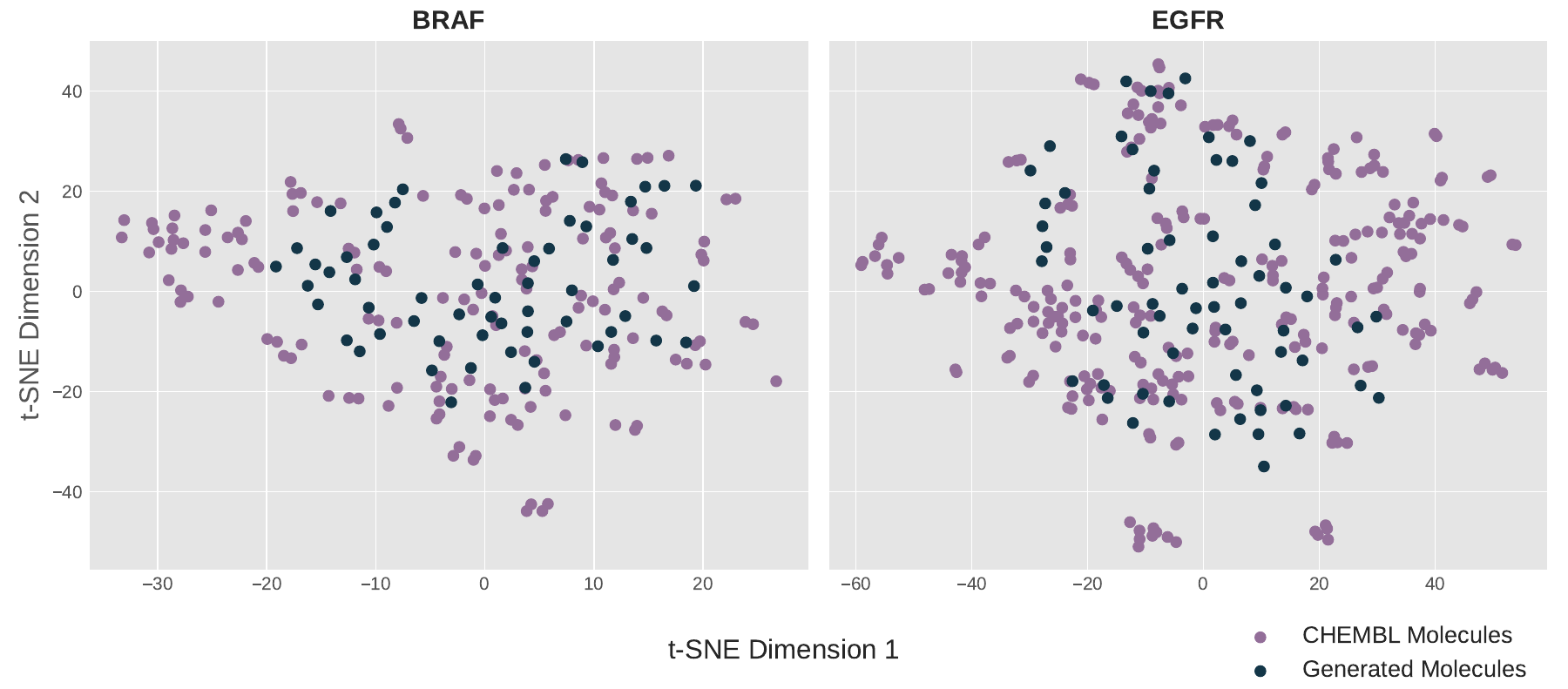}
\caption{T-SNE visualizations of fingerprint descriptors show that many generated compounds closely match or resemble reference molecules.}\label{tsne}
\end{figure}

\subsection{Assessing Molecular Novelty}

We assessed our methodology by inputting proteins into our RL fine-tuned model and examining the novelty of the generated molecules compared to the training data. This analysis aimed to determine whether the model merely memorized molecules. Our findings revealed that a mere 0.041\% of molecules were memorized from the training dataset. This underscores the model's capability to generate novel molecules.


\subsection{Investigation of Model-Generated Molecules}
To delve deeper into the outcomes produced by the model, we examined its outputs using a diverse range of proteins, such as human Bruton's tyrosine kinase (BTK), poly ADP-ribose polymerase (PARP), v-Raf murine sarcoma viral oncogene homologue B (BRAF), G-protein coupled receptor 6 (GPR6), and epidermal growth factor receptor (EGFR), for a comprehensive evaluation. Subsequently, we extracted a collection of Inhibitors known to interact with each protein from the ChEMBL database \cite{davies2015chembl} and performed Tanimoto coefficient \cite{bajusz2015tanimoto} similarity calculations between the generated compounds and reference compounds from ChEMBL database. Figure \ref{BigTable} visually represents a subset of generated and reference compounds and their corresponding Tanimoto Similarity scores, enhancing comprehensibility. Further visual representations are available in appendix \ref{A}. Notably, although the model was fine-tuned on protein amino acids of length 500, it effectively generated targeted compounds for proteins with longer sequences, such as BTK, BRAF, PARP, and EGFR. Moreover, we utilized the t-distributed Stochastic Neighbor Embedding (t-SNE) algorithm \cite{van2008visualizing} to visualize samples of the generated compounds' fingerprint descriptors and reference compounds' fingerprint descriptors in two dimensions, aiming to investigate whether the properties of the generated compounds aligned with the ChEMBL data. Figure \ref{tsne} illustrates the results for BRAF and EGFR, indicating that several generated compounds match or closely resemble reference molecules. Further visualizations are available in appendix \ref{B}.

\begin{table}
\centering
\resizebox{10.0cm}{!}{%
\begin{tabular}{ccccccc}

\toprule

\multirow{2}{*}{\textbf{Method}} & \multirow{2}{*}{\textbf{Protein}} & \multicolumn{4}{c}{\textbf{Metrics}} \\
&& Novel & Unique & Diversity & Filters \\

\midrule
AAE \cite{makhzani2015adversarial}& - & 0.793  & 1.0 & 0.855 & 0.996  \\
\midrule
JTN-VAE \cite{jin2018junction} & - & 0.914  &1.0 & 0.855 & 0.976  \\
\midrule
VAE \cite{kingma2013auto} & - & 0.694  &1.0 & 0.855 & 0.997  \\
\midrule
CharRNN \cite{segler2018generating} & - & 0.8419  & 1.0 & 0.856 & 0.994  \\
\midrule
latentGAN \cite{prykhodko2019novo} & - &  0.949 & 1.0 & 0.856 & 0.973 \\
\midrule
FSM-DDTR \cite{monteiro2023fsm} & - & 0.9596  & 0.998 & 0.871 & -  \\
\midrule

\multirow{4}{*}{Zhang et al. \cite{zhang2023universal}} 
& BTK & 0.990 & - & 0.674 &0.308 \\

& BRAF & \textbf{0.989} & - & 0.666 & 0.413 \\
 
    & EGFR & 0.979 & - & 0.702 & \textbf{0.793} \\

&PARP& 0.992 & - & \textbf{0.979} &0.398 \\

\midrule

\multirow{7}{*}{Proposed Method} 
& BTK & \textbf{1.0} & \textbf{1.0} & \textbf{0.853} & \textbf{0.666} \\

& BRAF & 0.982 & \textbf{1.0} & \textbf{0.858} & \textbf{0.672} \\
 
& EGFR & \textbf{1.0} & \textbf{1.0} & \textbf{0.853} & 0.630 \\

&PARP& \textbf{1.0} & \textbf{1.0} & 0.8481 &\textbf{0.614} \\

&GPR6& \textbf{1.0} & \textbf{1.0} & \textbf{0.849} & \textbf{0.638} \\

& Different-Proteins@1k & 1.0 & 0.993 & 0.835 & 0.84 \\

& Different-Proteins@6k & 1.0 & 0.962 & 0.834 & 0.835 \\
 
\bottomrule
\end{tabular}%
}
\vspace{0.3cm} 
\caption{Evaluation of the
molecular generation models utilizing MOSES.}
\vspace{-15pt}
\label{tab:comparisonTable}
\end{table}

\subsection{Performance Analysis}
We evaluated our approach against alternative molecular generation methods using Molecular Sets (MOSES) \cite{10.3389/fphar.2020.565644}, a benchmarking platform designed to facilitate drug discovery research. MOSES provides a comprehensive set of metrics for assessing the quality and diversity of generated molecules. Our analysis focused on several criteria: novelty, uniqueness, filters, and internal diversity. First, we performed the evaluation using molecules generated from five selected proteins as input to the model. Then, we extended the analysis by inputting a diverse set of 1,359 proteins rather than focusing on specific proteins and assessed the metrics at two scales: 1,000 molecules and 6,587 molecules. The results presented in Table \ref{tab:comparisonTable} demonstrate the efficacy of the proposed method across a range of protein profiles.

\section{Conclusion and Futurework}
This paper presents a targeted de novo drug design method that employs a Reinforcement Learning (RL) policy to generate compounds specifically designed to interact with proteins. For any given target protein, the method produces a tailored molecular compound. Through extensive evaluations across diverse scenarios, we demonstrate the effectiveness of our approach. Our framework efficiently generates protein-specific chemical compounds while ensuring favourable chemical properties, such as Molecular Weight (MW), Quantitative Estimation of Drug-likeness (QED), and Octanol-Water Partition Coefficient (logP). Future work will focus on exploring larger generative models and datasets and experimenting with alternative reward functions. This will include metrics such as uniqueness, filters, and internal diversity from the MOSES benchmark, aimed at further enhancing performance and testing the ability to refine the learning policy.




\bibliographystyle{plain}
\bibliography{custom}

\clearpage

\onecolumn
\appendix

\section{Exploring Similarity of Generated Molecules and Protein Inhibitors}\label{A}
\begin{figure*}[ht]
\centering
\includegraphics[width=0.7\textwidth]{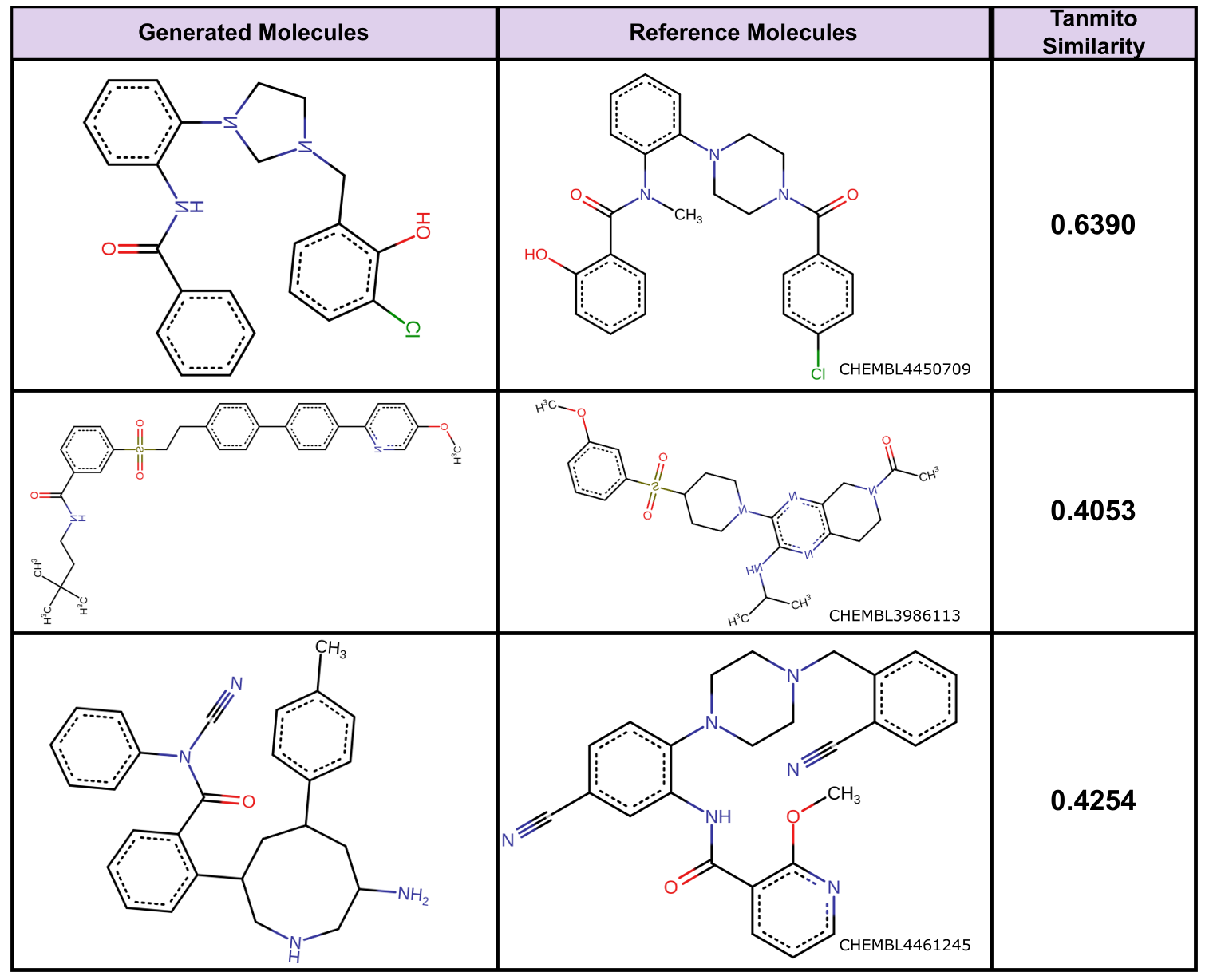}
\caption{Examples of molecules generated by inputting the G-protein coupled receptor 6 (GPR6) protein to the model, alongside samples of protein inhibitors, with the Tanimoto similarity ($TS$) measured between the generated and inhibitor compounds.}
\label{GPR6}
\end{figure*}

\begin{figure*}[ht]
\centering
\includegraphics[width=0.7\textwidth]{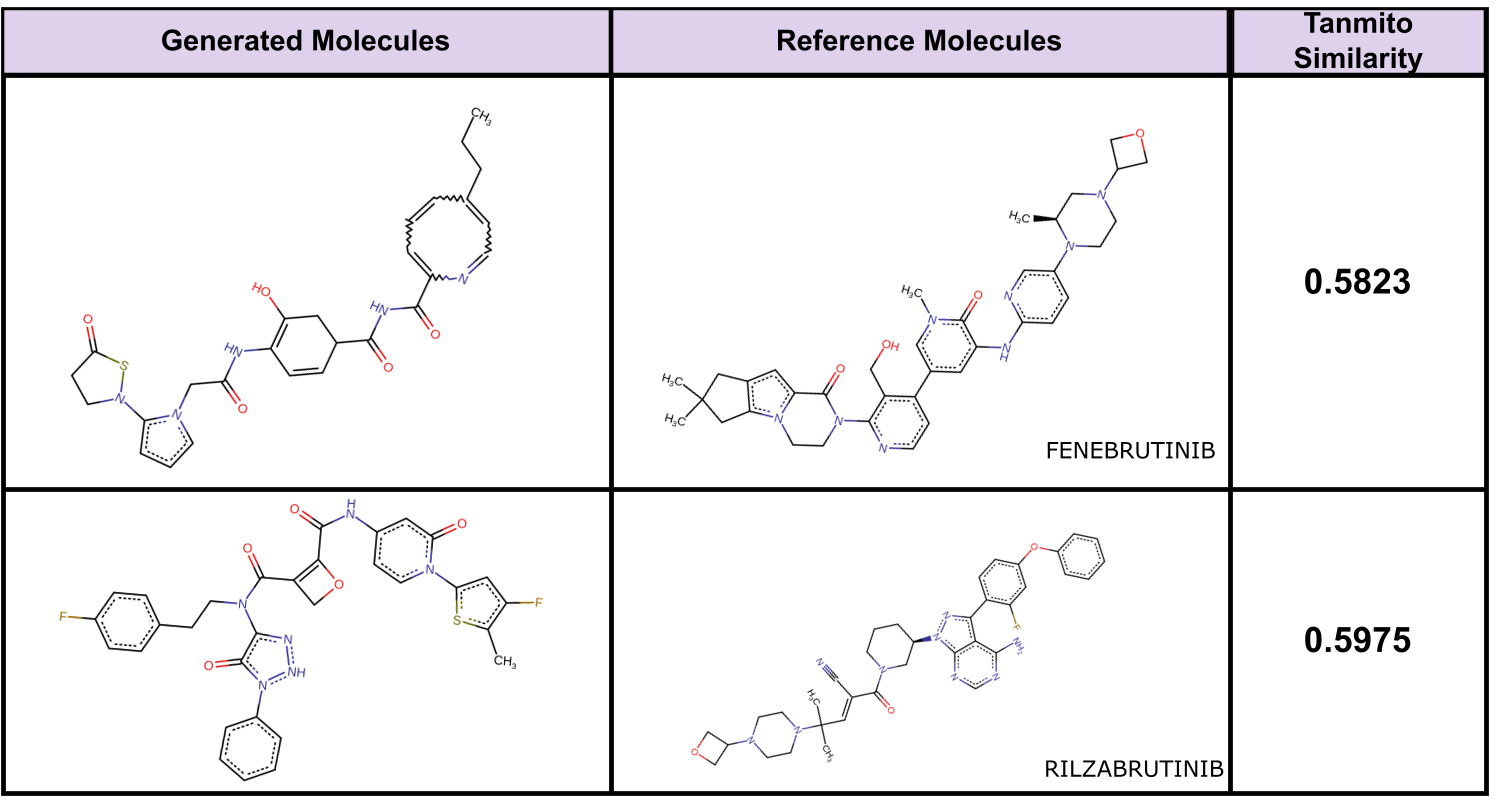}
\caption{Examples of molecules generated by inputting the Bruton's tyrosine kinase (BTK) protein to the model, alongside samples of protein inhibitors, with the Tanimoto similarity ($TS$) measured between the generated and inhibitor compounds.}
\label{BTK}
\end{figure*}

\begin{figure*}[ht]
\centering
\includegraphics[width=0.7\textwidth]{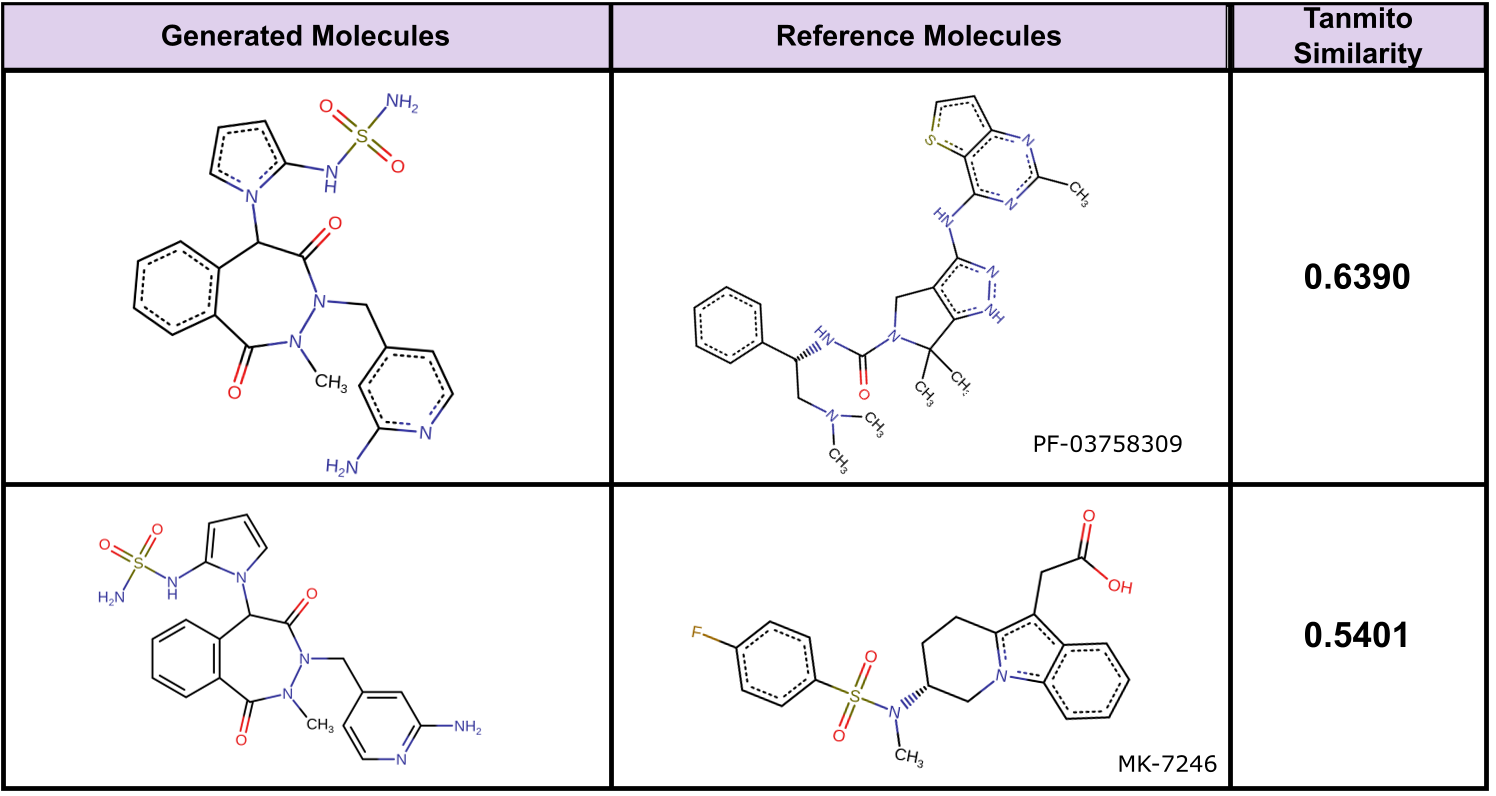}
\caption{Examples of molecules generated by inputting the v-Raf murine sarcoma viral oncogene homologue B (BRAF) protein to the model, alongside samples of protein inhibitors, with the Tanimoto similarity ($TS$) measured between the generated and inhibitor compounds.}
\label{BRAF}
\end{figure*}

\begin{figure*}[ht]
\centering
\includegraphics[width=0.7\textwidth]{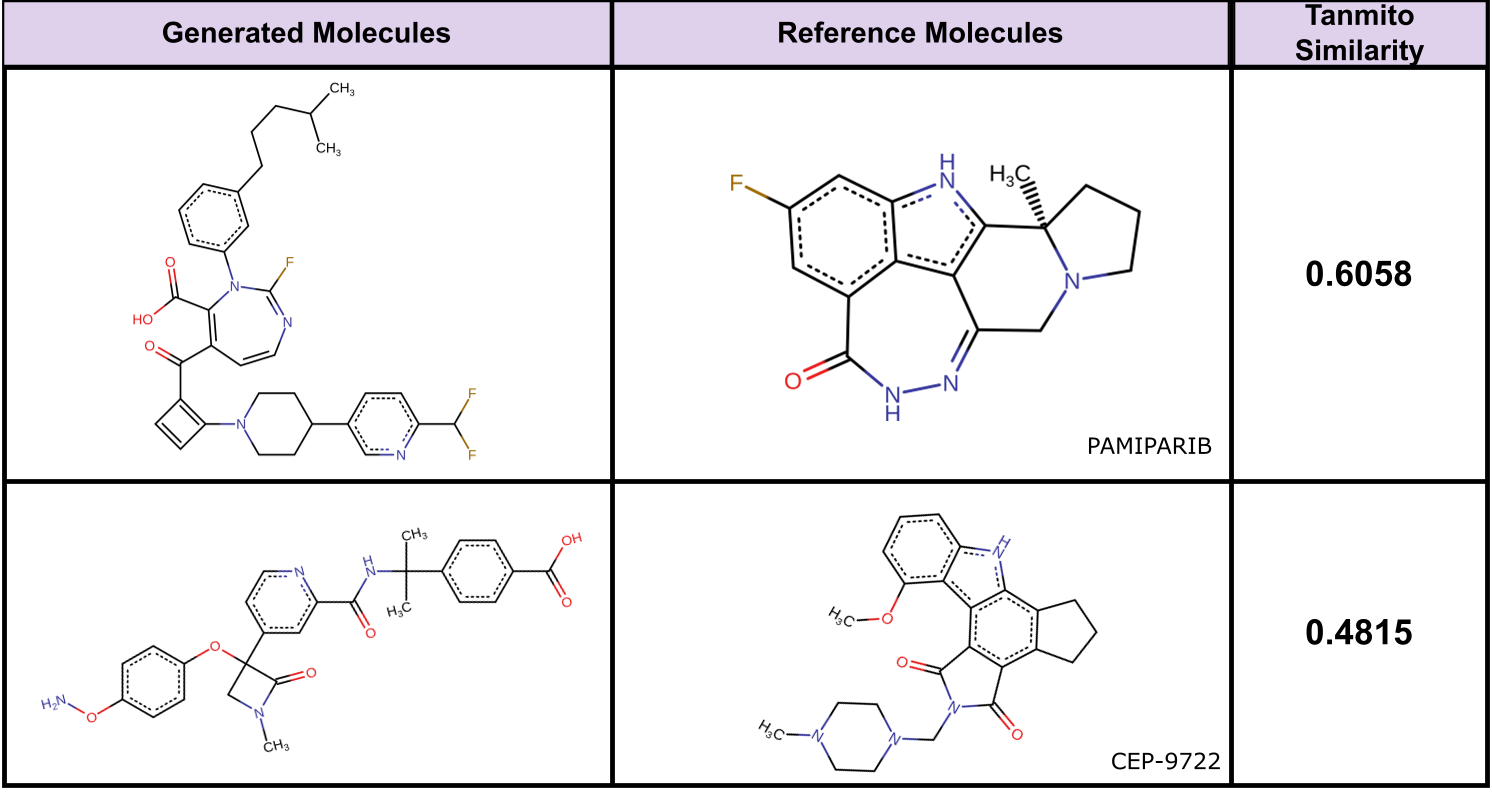}
\caption{Examples of molecules generated by inputting the poly ADP-ribose polymerase (PARP) protein to the model, alongside samples of protein inhibitors, with the Tanimoto similarity ($TS$) measured between the generated and inhibitor compounds.}
\label{PARP}
\end{figure*}

\begin{figure*}[ht]
\centering
\includegraphics[width=0.7\textwidth]{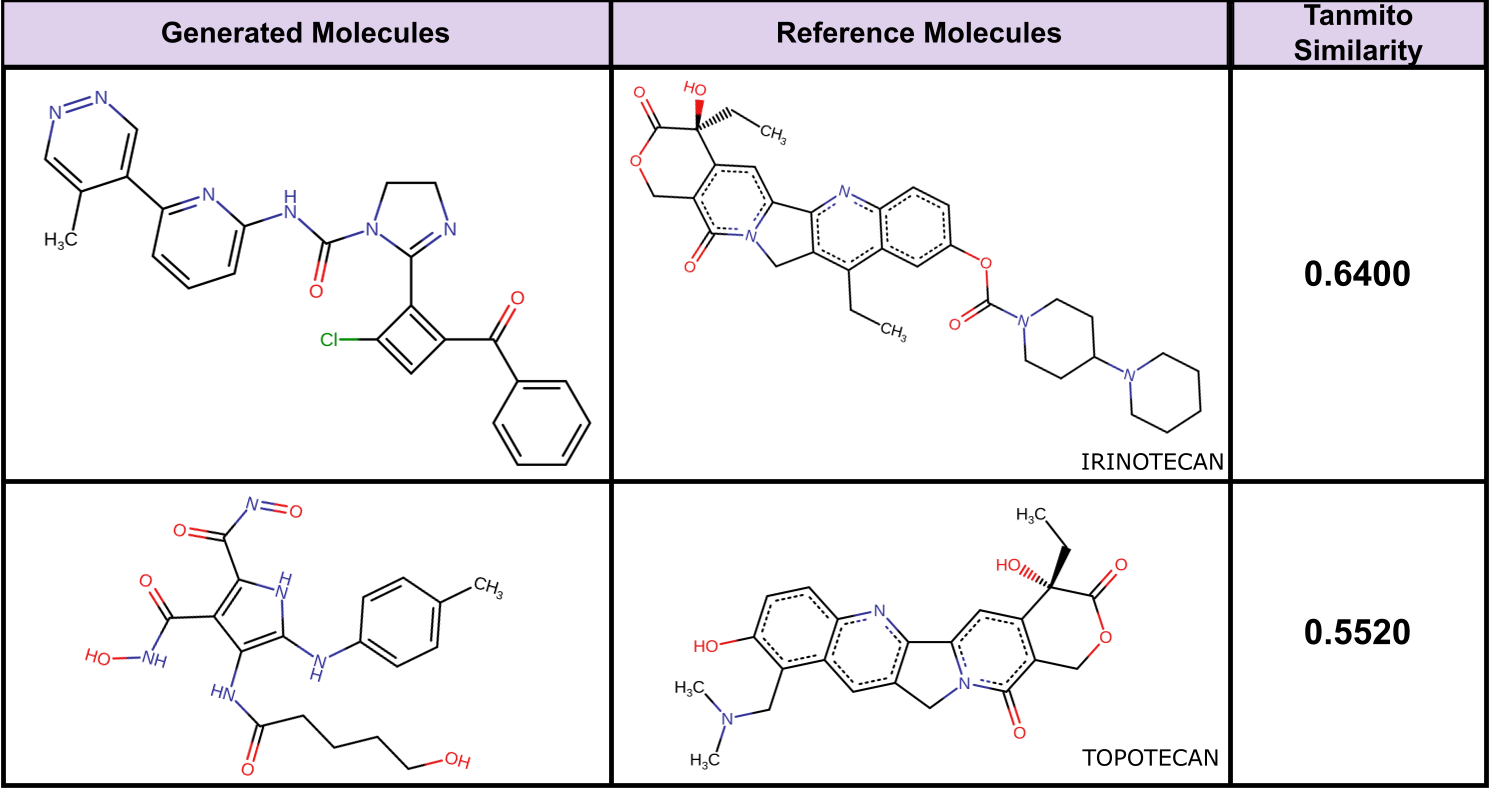}
\caption{Examples of molecules generated by inputting the epidermal growth factor receptor (EGFR) protein to the model, alongside samples of protein inhibitors, with the Tanimoto similarity ($TS$) measured between the generated and inhibitor compounds.}
\label{EGFR}
\end{figure*}

We showcase further instances of molecules generated by inputting different proteins into the model. These molecules are matched with protein inhibitors sourced from the ChEMBL database. Compared to the inhibitors, their similarity is assessed using the Tanimoto similarity (TS) metric. These instances are visually represented in Figures \ref{GPR6} through \ref{EGFR}.

\clearpage
\section{Visualizing Molecular Similarity with t-SNE}\label{B}
\begin{figure*}[ht]
\centering
\includegraphics[width=0.6\textwidth]{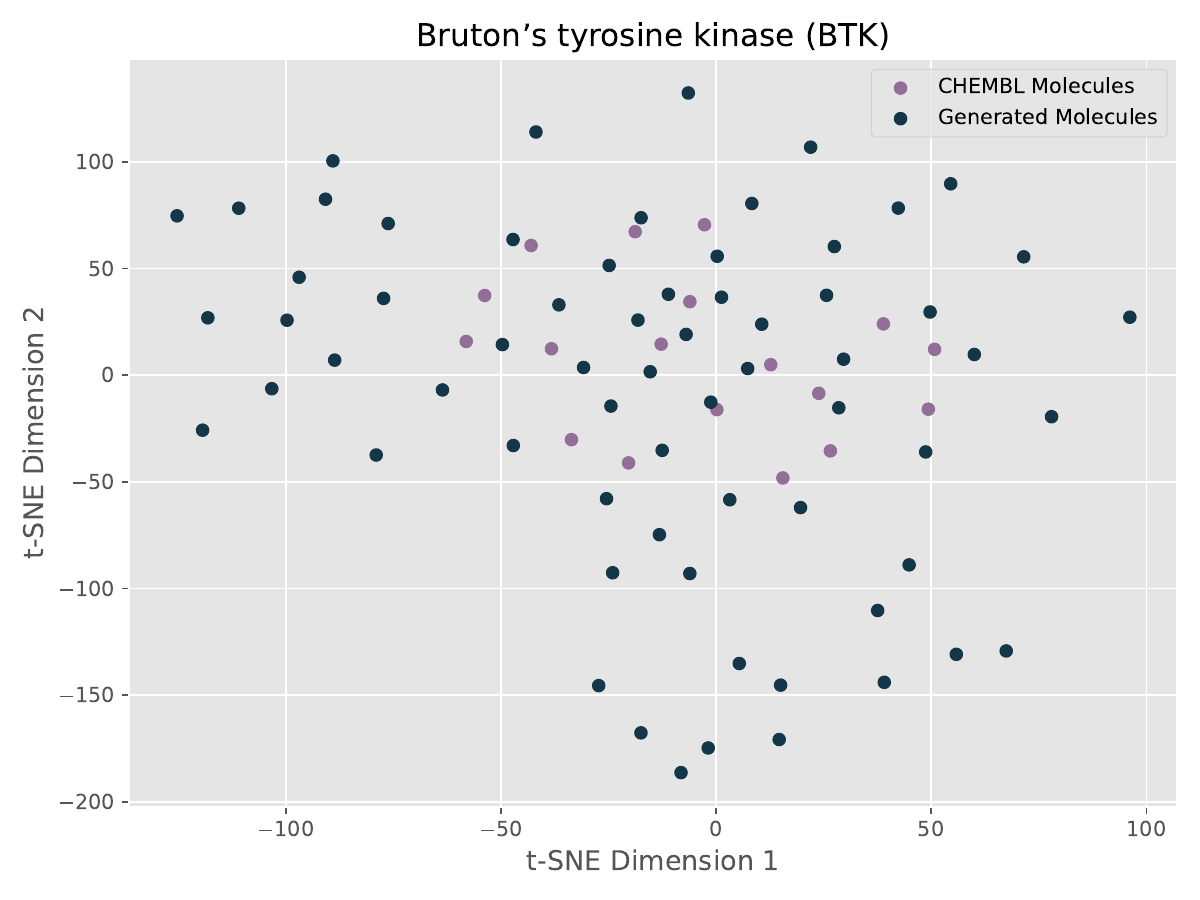}
\caption{Visual representations using t-SNE projections illustrating the fingerprint descriptors of both generated and reference molecules for BTK.}
\label{tsneBTK}
\end{figure*}

\begin{figure*}[ht]
\centering
\includegraphics[width=0.6\textwidth]{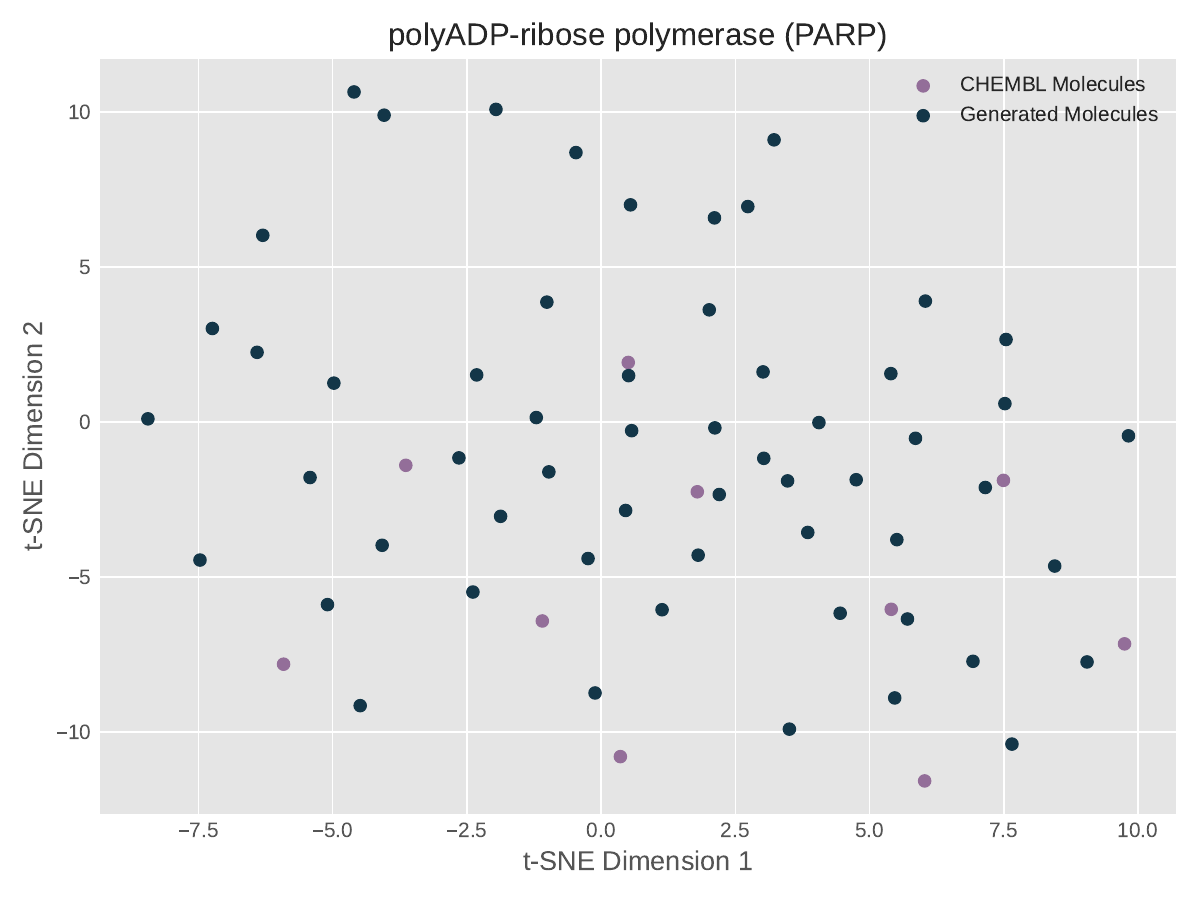}
\caption{Visual representations using t-SNE projections illustrating the fingerprint descriptors of both generated and reference molecules for PARP.}
\label{tsnePARP}
\end{figure*}

\begin{figure*}[ht]
\centering
\includegraphics[width=0.6\textwidth]{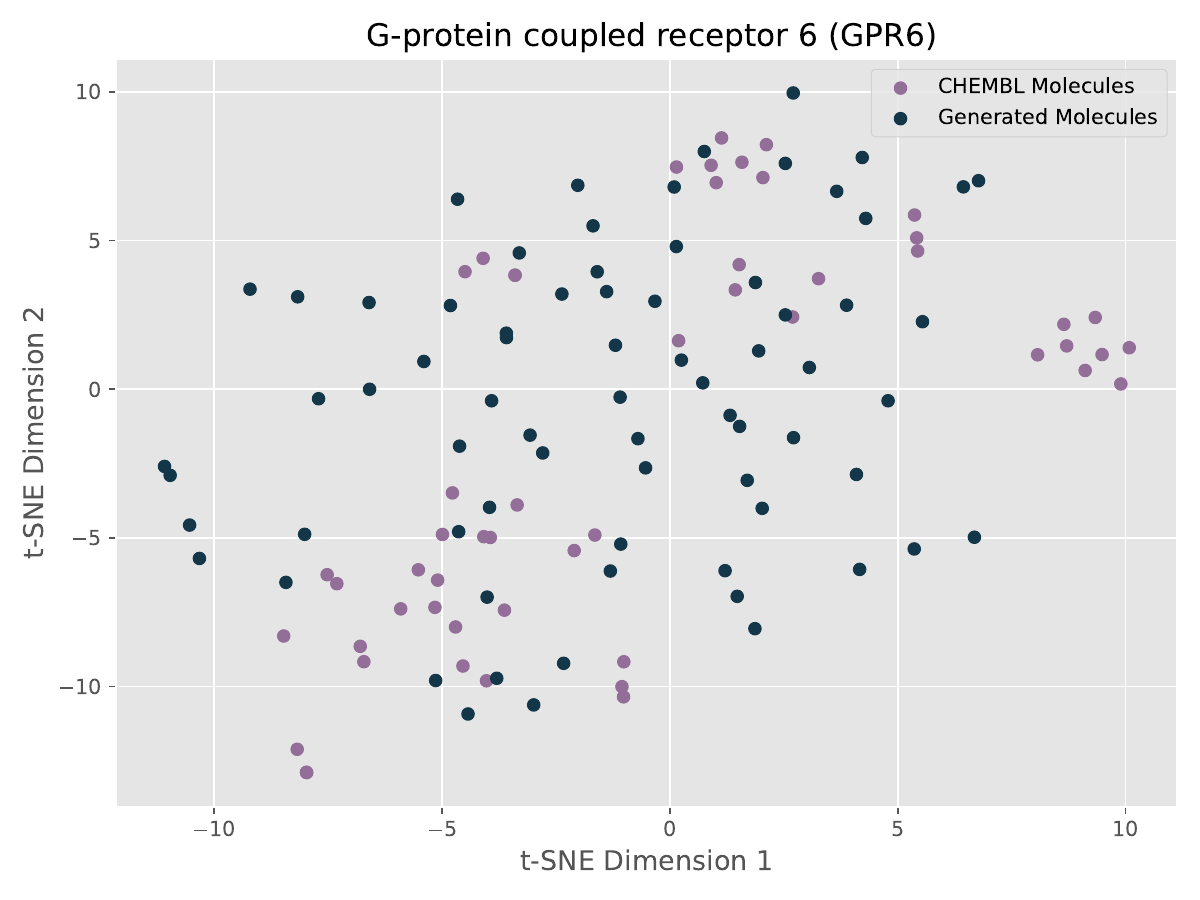}
\caption{Visual representations using t-SNE projections illustrating the fingerprint descriptors of both generated and reference molecules for GPR6.}
\label{tsneGPR6}
\end{figure*}

We employed molecular fingerprint descriptors to visually represent the generated molecules and protein inhibitors utilizing the Stochastic Neighbor Embedding (t-SNE) algorithm. This visualization highlights their similarity, further affirming the efficacy of our proposed method. Figures \ref{tsneBTK} to \ref{tsneGPR6} elucidate these findings.

\clearpage

\end{document}